\documentclass[manuscript]{aastex631} 
\usepackage{graphicx} 

\begin{document}

\title{Around the Spindle galaxy: the dark halo mass of NGC3115.}

\author{I. D. Karachentsev}
\affiliation{Special Astrophysical Observatory, the Russian Academy of Sciences, Nizhnij Arkhyz, KChR, 369167, Russia}

\author{L. N. Makarova}
\affiliation{Special Astrophysical Observatory, the Russian Academy of Sciences, Nizhnij Arkhyz, KChR, 369167, Russia}

\author{G. S. Anand}
\affiliation{Space Telescope Science Institute, 3700 San Martin Drive, Baltimore, MD 21218, USA}

\author{R. B. Tully}
\affiliation{Institute for Astronomy, University of Hawaii, 2680 Woodlawn Drive, Honolulu, HI 96822, USA}

\begin{abstract}
We report observations of five dwarf galaxies in the vicinity of the luminous S0 galaxy NGC\,3115 performed with the 
Advanced Camera for Surveys on the Hubble Space Telescope.  Their distances determined via the Tip of the Red Giant 
Branch are: 10.05 Mpc (UGCA\,193), 9.95 Mpc (KKSG 17), 10.13 Mpc  (2MASX-J0957-0915), 10.42 Mpc (2dFGRS-TGN218Z179) 
and 11.01 Mpc (KKSG\,19). With their typical distance error of about 0.75 Mpc all the five dwarfs are consistent to 
be true satellites of the host galaxy NGC\,3115 (10.2$\pm$0.2 Mpc). Using the DESI Legacy Imaging Surveys we also found 5 
new probable dwarf satellites of NGC\,3115, as well as 4 new probable members of the neighboring group around NGC\,3521 
situated 3 Mpc away from the NGC\,3115 group. Based on the radial velocities and projected separations of 10 dwarf 
companions, we derived the total (orbital) mass of NGC\,3115 to be (4.89$\pm$1.48) $10^{12}$ $M_{\odot}$. The ratio 
of the total mass-to-K-luminosity of NGC\,3115 is (50$\pm$15) $M_\odot/L_\odot$, which is typical for the early-type 
luminous galaxies.
\end{abstract}

\keywords{galaxies: distances and redshifts - galaxies: dwarf}

\section{Introduction.}
According to modern concepts, the bulk of the matter of galaxies is contained in their dark halos, the nature of which 
remains unclear.  The fundamental parameter of a galaxy is the ratio between its total mass, $M_T$, and
stellar mass, $M_*$.  The  dimensionless quantity $M_T/M_*$ characterizes  the history of star formation in the galaxy, 
as well as its present dynamic status.  In recent years, many publications have appeared in which the magnitude of 
$M_T/M_*$ is modeled depending on the luminosity or mass of the galaxy (Sales et al. 2013; Moster et al. 2013; 
Wechsler \& Tinker 2018; Santos-Santos et al. 2021). As shown by theoretical calculations and observational results, 
the minimum ratio $M_T/M_*$ occurs for galaxies of the Milky Way type with $M_*/M_{\odot}
\sim11$ dex, increasing both towards more massive galaxies and towards dwarf systems (Kourkchi \& Tully 2017; Lapi 
et al. 2018; Posti \& Fall 2021).
 
 In addition to the dependence on the mass (luminosity) of a galaxy, the ratio $M_T/M_*$ also depends on its 
morphological type. Spiral galaxies of type Milky Way and M~31 have an $M_T/M_* \sim 25$, while early-type galaxies 
(ETG)  with dominant bulges are characterized by an $M_T/M_*$ ratio that is 2--3 times higher (Karachentseva et al. 
2011; Mandelbaum et al. 2016; Bilicki et al. 2021).  This difference may be due to the fact that star formation in 
disk-dominated galaxies occurs fairly evenly throughout the entire cosmological scale
of  13.8 Gyr, while the main (violent) formation of stars in E and S0 galaxies ended in an early epoch, almost 
ceasing in the last 10 Gyr. 
  At the same time, early-type galaxies are not inferior to late-type galaxies in terms of integral luminosity and 
stellar mass.  For example, in the Local Volume, limited by the distance $D < 11$ Mpc, the five brightest relatively 
isolated galaxies of the S0 type (NGC 4594, NGC 3115, NGC 1291, NGC 5128 and NGC 2784) have an average K-band 
luminosity  $\langle \log(L_K/L_{\odot})\rangle = 10.99\pm0.09$, while the five brightest spiral galaxies of the 
Sc-Scd type
(NGC 6946, NGC 253, NGC 5194, NGC 5236 and M~101) have an average luminosity of 
$\langle \log(L_K/L_{\odot})\rangle = 10.92\pm0.04$
(Karachentsev \& Kashibadze 2021).
  Obviously, the $M_T/M_*$ ratio in galaxies is influenced not only by the history of star formation in them, 
but also by the repeated process of merging of the galaxy with its surrounding neighbors. Being in a denser 
environment, E and S0 galaxies are likely to experience more frequent mergers and, therefore, turn out to be among 
the leaders in terms of integral luminosity.
  The total (dynamic) mass of E and S0 galaxies is usually determined from the dispersion of the line-of-sight 
velocities of stars, globular clusters or planetary nebulae (Capaccioli et al. 1993; Peng et al. 2004; Guerou et 
al. 2016; Bilek et al. 2019). These estimates correspond to a linear scale of $\sim$(20--50) kpc  and obviously 
underestimate the total mass of the galaxy within the virial radius of its halo, $R_{vir} \sim 350$ kpc.  To 
determine $M_T$ for isolated E and S0 galaxies, data on radial velocities and projected separations of small 
satellites of these galaxies are most suitable.   Of particular interest here are the ETG galaxies of the Local 
Volume, where the observational data on accurate distances and velocities of galaxies  are most complete in 
comparison with distant volumes. At present, fairly reliable estimates of the total mass have been made for only 
two ETG galaxies of the Local Volume: NGC~5128 (Centaurus A) and NGC~4594 (Sombrero) (Karachentsev et al. 2020; 
Karachentsev \& Kashibadze 2021).  In this article, we measure the distances to 5 satellites of the S0 galaxy 
NGC~3115 (Spindle galaxy) and determine the total mass of this host galaxy.
 
 In section 2, we present images of 5 dwarf galaxies in the vicinity of NGC~3115, obtained with the Hubble Space 
Telescope (HST),  and estimate the distances of these galaxies via the Tip of Red Giant Branch (TRGB) in
color-magnitude diagrams (CMD).  In section 3, we report on the detection of five new candidates for satellites of 
NGC~3115 and give a summary of the parameters of 27 putative satellites of NGC~3115, whose radial velocities and 
projected separations determine the orbital (total) mass of the central galaxy, $M_T$. Section 4 is devoted to the 
group of dwarf satellites around the Sbc galaxy NGC~3521, adjacent to the group around NGC~3115. The dark halo of 
galaxy NGC~3521 is distinguished by an abnormally low $M_T/M_*$ ratio.  Section 5
presents a summary of mass estimates for 5 isolated high-luminosity ETC-galaxies in the Local Volume. Our conclusions 
are summarized in section 6.

  \section{HST observations and  new TRGB distances.}
  One hundred and fifty target galaxies were selected by us for observations with the ACS camera on the HST within 
the SNAP-15922 program  "Every Known Nearby Galaxy" (PI R.B. Tully).  Of these, 80 galaxies, mainly dwarf systems, 
were observed in the filters F606W and F814W with exposures of 760 sec in each filter  during 2019-2020. The five
observed galaxies are located in the vicinity of the bright ( B = 9.9 mag) lenticular galaxy  NGC~3115, which has a 
heliocentric radial velocity  $V_h = (681\pm6)$ km s$^{-1}$ (Wegner et al. 2003). The distance to it, 
$D = (9.68\pm0.40)$ Mpc, was determined by Tonry et al. (2001) via surface brightness fluctuations (SBF). Later 
Peacock et al (2015) specified its distance, $D = (10.2\pm0.2)$ Mpc, using the TRGB method. One of the NGC~3115 
neighbors we observed, UGCA~193 = PGC~029086, is a spiral Sd galaxy seen edge-on. Its angular Holmberg diameter, 
4.3 arcmin, exceeds the size of the ASC field-of-view.  The remaining 4 objects: KKSG~17=MCG-01-26-011=PGC~029033, 
2MASX-J09570887-0915487=PGC~154449, 2dFGRS-TGN218Z179=PGC~1099440 and KKSG~19= PGC~3097700 are irregular (dIrr) or 
blue compact dwarf (BCD) galaxies with angular sizes of 0.7 - 0.9 arcmin. Their ACS images, created with the F606W 
(blue), F814W (red), and the average of the two images (green) are shown in the Figure 1.

\begin{figure}
    \centering
    \includegraphics[width=0.3\textwidth]{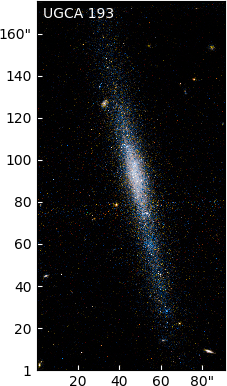}
    \includegraphics[width=0.5\textwidth]{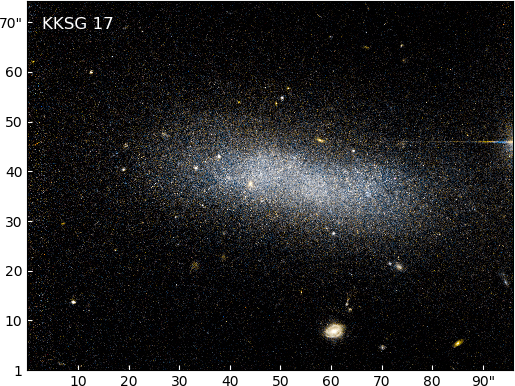}
    \includegraphics[width=0.5\textwidth]{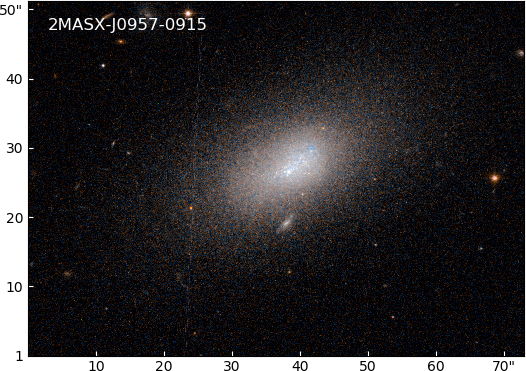}
    \includegraphics[width=0.4\textwidth]{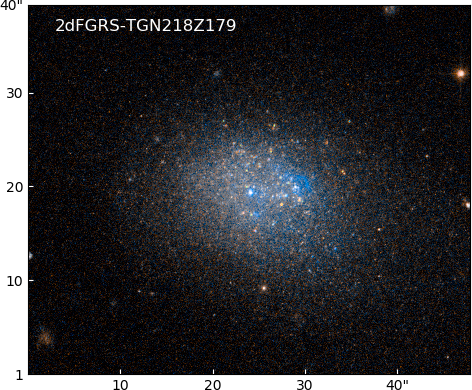}
    \includegraphics[width=0.4\textwidth]{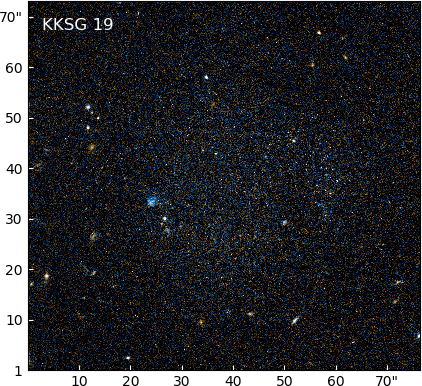}
    \caption{Two-color HST/ACS images of five dwarf galaxies around NGC~3115: UGCA~193, KKSG~17, 2MASX-J0957-0915, 2dFGRS-TGN218Z179 
and KKSG~19. North is up and east is left.}
\end{figure}
 
 The obtained images were processed by the standard HST pipeline (Dolphin 2000, 2016) with the recommended DOLPHOT 
parameters: ($Crowd_{\rm F606W} + Crowd_{\rm F814W}) < 0.8$ and ($Sharp_{\rm F606W} + Sharp_{\rm F814W})^2 < 0.075$. 
Stars with a signal-to-noise ratio $S/N \geq 3$ in {\rm F606W} and
$S/N \geq 4$ in {\rm F814W} filter were selected. Since each  of
the 5 galaxies occupied a small area of the ACS field of view, we  excluded from the analysis objects located far 
beyond the optical boundaries of galaxies.  In case of UGCA 193, we also excluded objects in the central part of 
the galactic disk with  extremely high stellar density.
 
\begin{figure}
    \centering
    \includegraphics[width=0.3\textwidth]{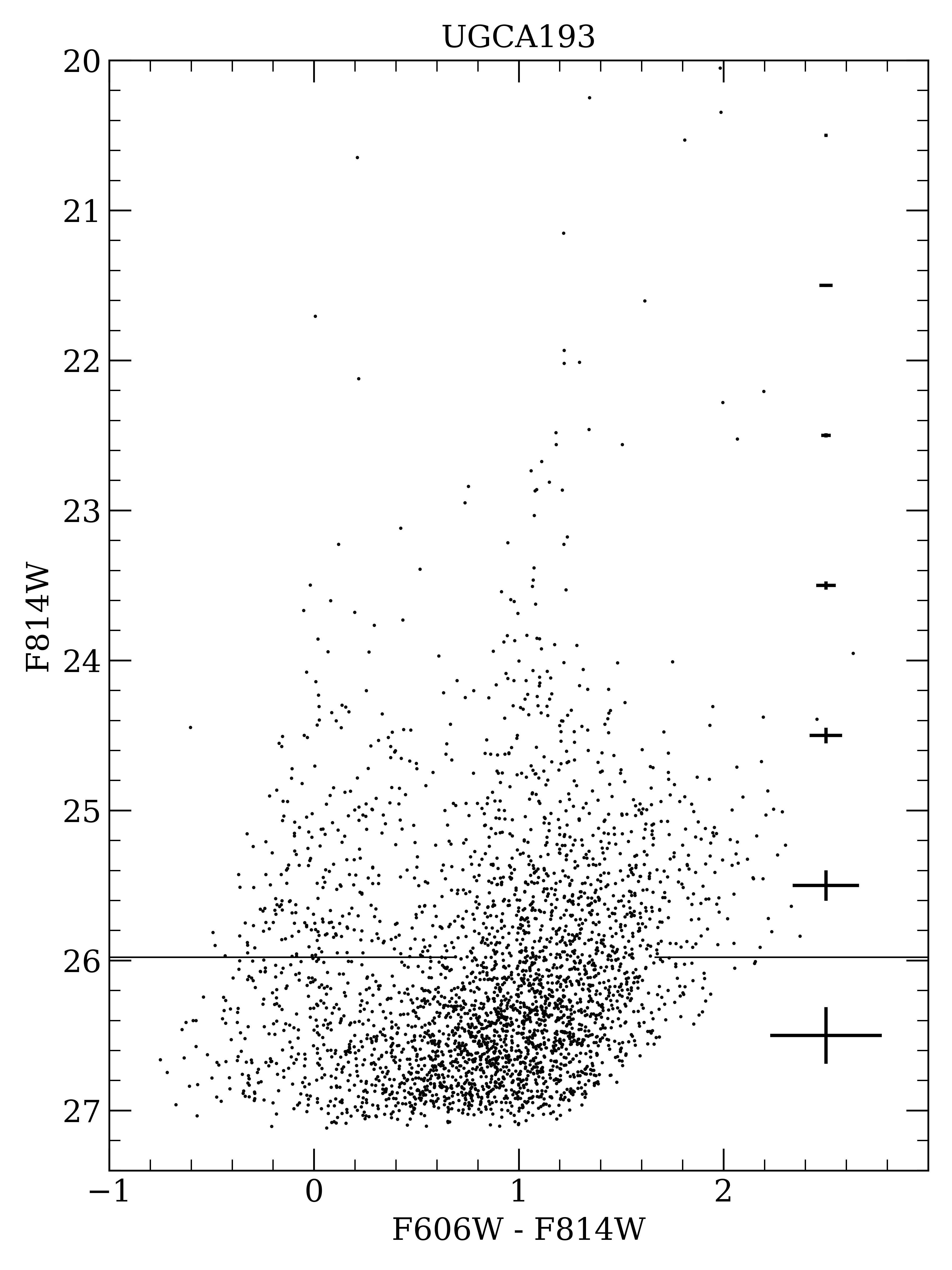}
    \includegraphics[width=0.3\textwidth]{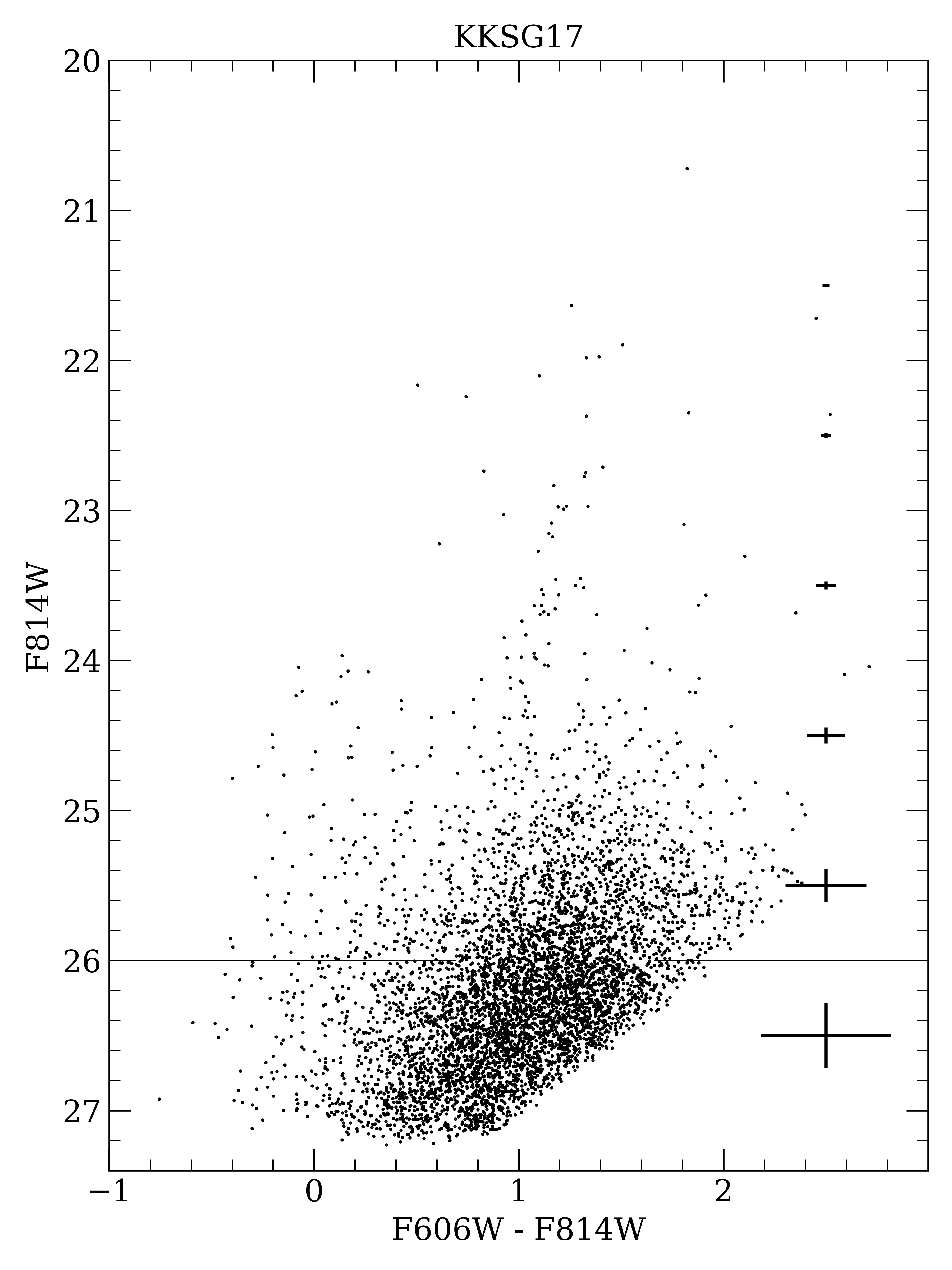}
    \includegraphics[width=0.3\textwidth]{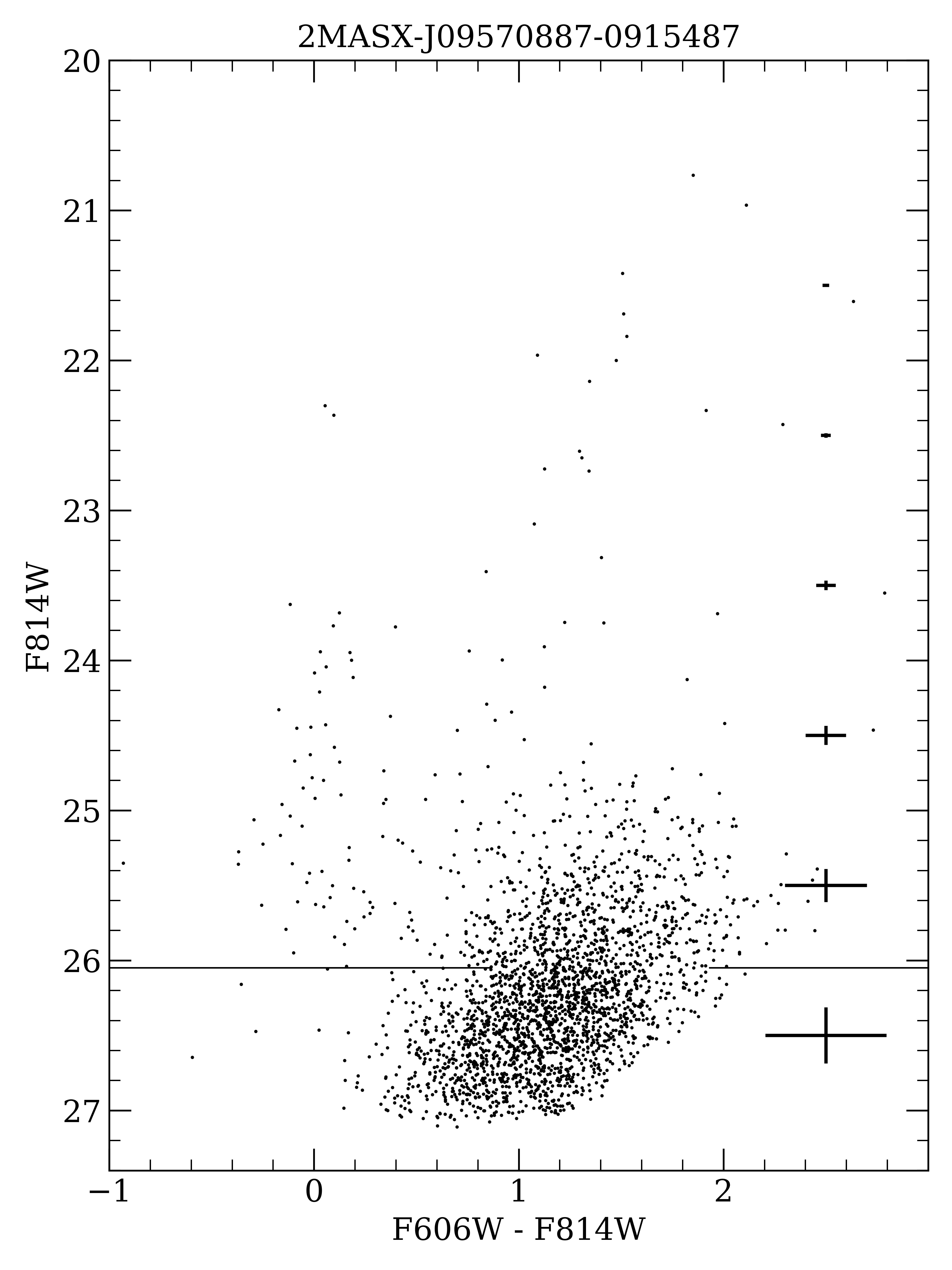}
    \includegraphics[width=0.3\textwidth]{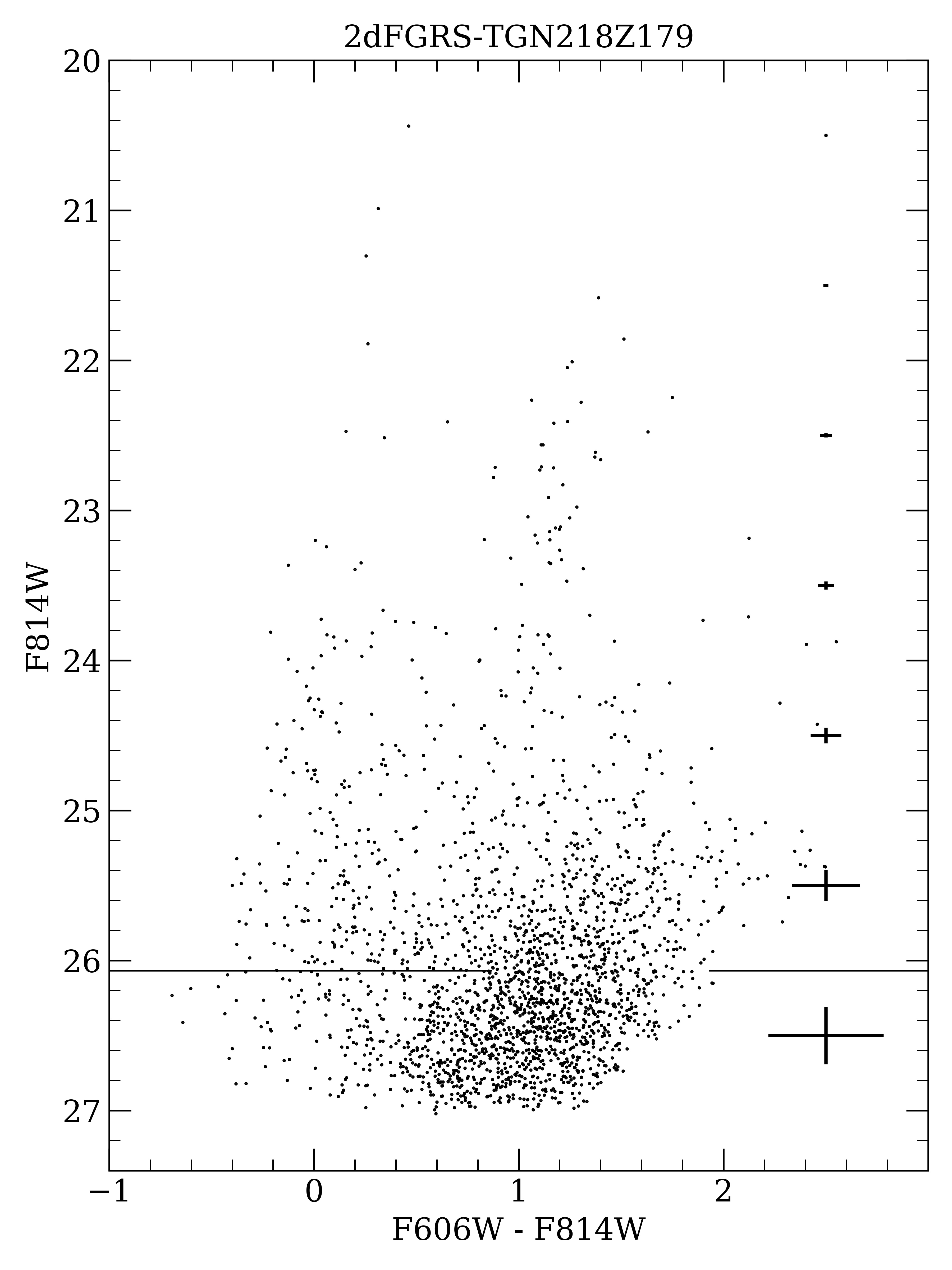}
    \includegraphics[width=0.3\textwidth]{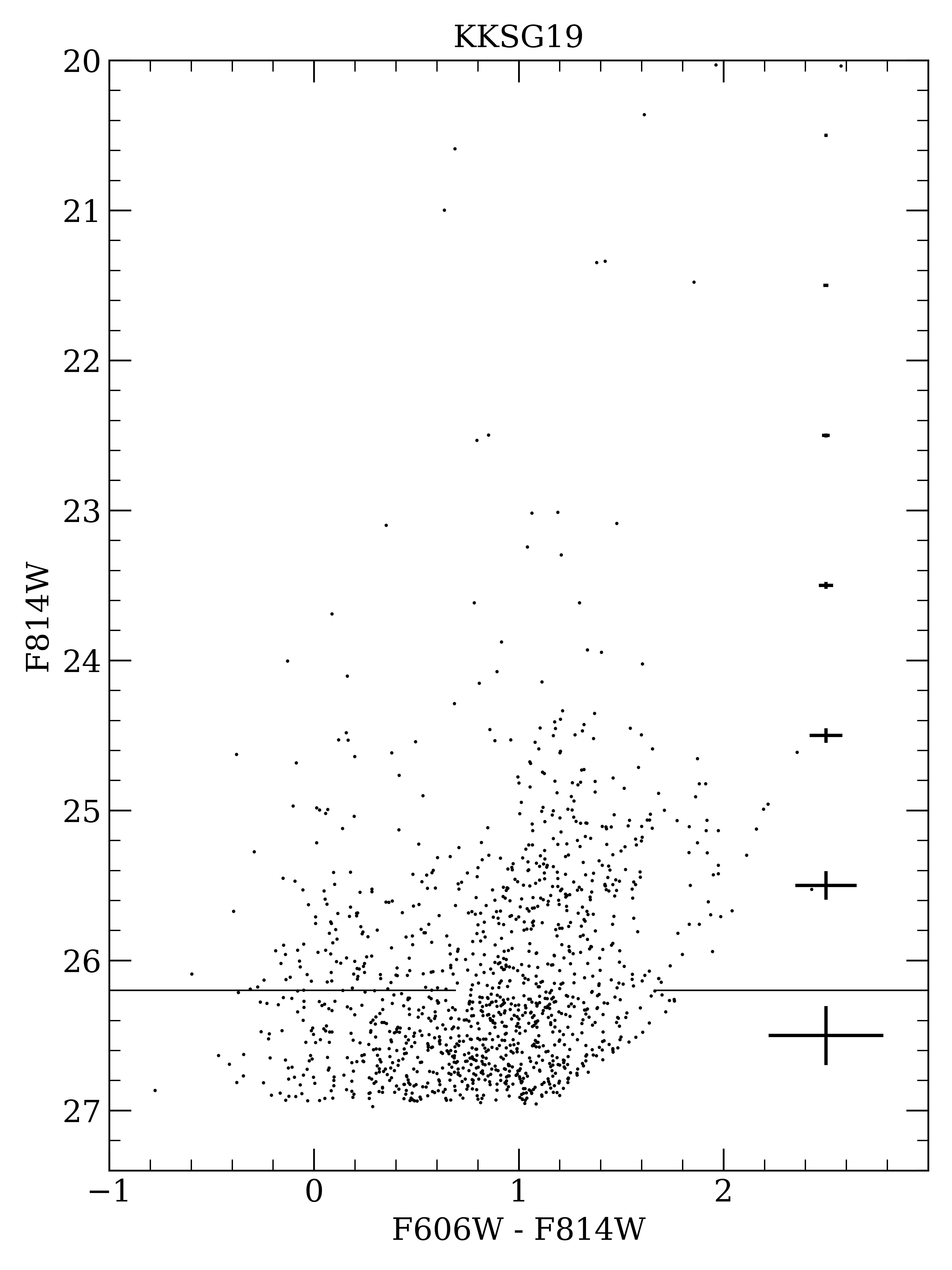}
    \caption{Color-magnitude diagrams of the observed galaxies. The TRGB positions are indicated by the horizontal 
lines. Photometric errors are shown by the bars at the right in the CMDs.}
\end{figure} 
 
 The resulting CMDs in F606W-F814W vs. F814W  are presented in five panels of Figure 2.  The diagrams show 
the stars of the young and old populations of galaxies. The position of the Tip of the Red Giant Branch was 
determined by the discontinuity on the luminosity function of stars within the colors $[0.5 < F606W -F814W < 1.5]$ 
in a maximum likelihood analysis described by Makarov et al. (2006) and updated by Wu et al. (2014). 
In this method, the luminosity functions of the Red Giant Branch stars and Asymptotic Giant Branch populations
are fitted by a broken power law, with the break signifying the location of the TRGB. The TRGB positions are marked on the CMDs 
with horizontal lines. Artificial stars were created and recovered using the same DOLPHOT parameters to estimate the 
photometric errors. The results are given in Table 1, where $F814W_{\rm TRGB}$ means the position of the TRGB, 
$A_{814}$ - interstellar extinction in the F814W filter according to Schlafly \& Finkbeiner (2011), $M_{\rm TRGB}$ 
--- the zero-point calibration of the absolute magnitude of the TRGB following Rizzi et al. (2007),
DM and D are the distance modulus (mag) and the distance (Mpc). The typical error in determining the TRGB is
about 0.15 mag, which corresponds to an error in the distance of $\sim$0.75 Mpc. ( Some doubts, however, remain about 
KKSG~19, whose TRGB position is close to the CMD photometric limit). Within this error, all distances of 
the dwarf galaxies are in excellent agreement with the distance of the central galaxy itself $(10.2\pm0.2$ Mpc). 
This gives reason to consider them as physical satellites of the lenticular galaxy NGC~3115.

\begin{table}
  \caption{TRGB properties of the dwarf companions of NGC 3115.}
  \begin{tabular}{lcccrc}\hline

  Galaxy                  &      $F814W_{\rm TRGB}$ &    $A_{\rm 814}$  &  $M_{\rm TRGB}$   &  $DM$   &   $D$\\
\hline
                           &        mag            &   mag        &   mag   &     mag  &     Mpc\\
 \hline
 (1)&(2)&(3)&(4)&(5)&(6)\\
\hline
 UGCA 193                  &      25.98        &         0.060    &    $-$4.09   &      30.01    &     10.05 \\
 KKSG 17                   &      26.00        &         0.090    &    $-$4.08   &      29.99    &      9.95 \\
 2MASX-J0957-0915           &     26.05         &        0.102     &   $-$4.08    &     30.03     &    10.13 \\
 2dFGRS-TGN218Z179          &     26.07         &        0.066     &   $-$4.09    &     30.09     &    10.42\\
 KKSG 19                    &     26.20         &        0.104     &   $-$4.11    &     30.21     &    11.01\\
\hline
\end{tabular}
\end{table}

  \section{ Retinue of the Spindle galaxy.}
  Examining the SERC EJ zone of the Palomar Sky Survey, Karachentsev et al. (2000) found six low surface brightness 
dwarf galaxies  in the vicinity of NGC~3115. Some of them were noted earlier in the MCG catalog (Vorontsov-Veliaminov \& Arkhipova, 1963).
Observations in the 21 cm HI line (Huchtmeier et al. 2001) showed that 
the radial velocities of these objects are close to the radial velocity of NGC~3115. Surveys of redshifts of southern
 galaxies: 2dF (Colles et al. 2003), HIPASS (Doyle et al. 2005) and 6dF (Jones et al. 2009) have added 5 more dwarf
 galaxies to the supposed satellites of NGC~3115. Using the DESI Legacy Imaging Surveys  = DECaLS (Dey et al. 2019), 
Carlsten et al. (2021a) found 11 more probable dwarf satellites of NGC~3115. Their search area in this digital survey
 of the northern sky covered only part of the virial zone around NGC~3115. Therefore, we repeated the search for faint 
satellites of NGC~3115, expanding the search area to 20 square degrees.  As a result, we found all 11 dwarf systems 
noted by Carlsten et al. (2021a), and on top of that found 5 more candidates for faint satellites of the Spindle galaxy.  
Color images of these galaxies taken from the DECaLS survey (Dey et al. 2019) are shown in Fig.3. The images size is 
2'x2', north is up and east is left. Table 2 contains: equatorial coordinates of the galaxies, their major angular 
diameter in arcmin, axial ratio, morphological type, and apparent B- magnitude  estimated by eye from a comparison of the new
 putative satellites with other dwarfs, whose g- and r- magnitudes were measured by Carlsten et al. (2021b). According
to our estimates, the error in determining the visual B- magnitudes is about 0.25 mag.

\begin{figure}
    \centering
    \includegraphics[width=0.4\textwidth]{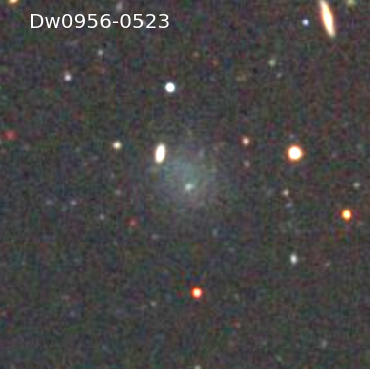}
    \includegraphics[width=0.4\textwidth]{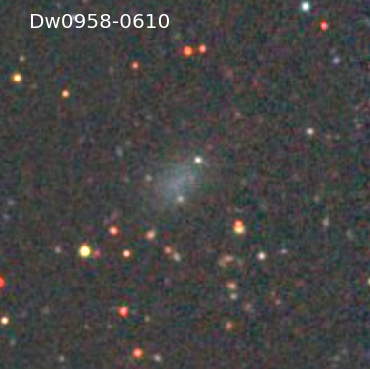}
    \includegraphics[width=0.4\textwidth]{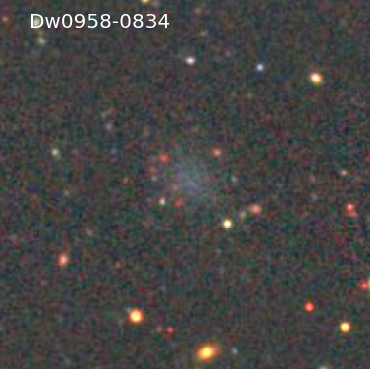}
    \includegraphics[width=0.4\textwidth]{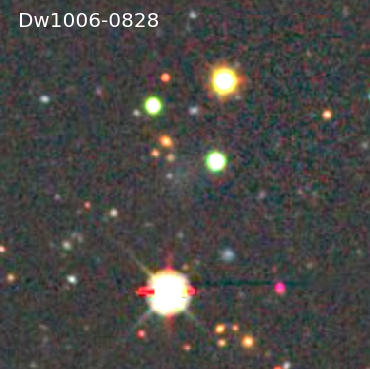}
    \includegraphics[width=0.4\textwidth]{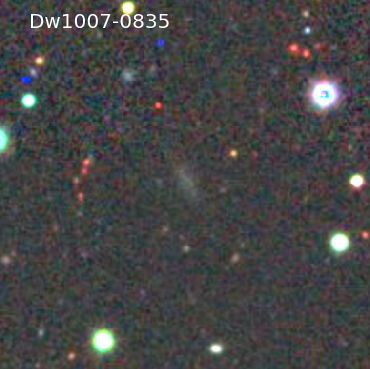}
    \caption{Color-composed images of five new probable satellites of NGC~3115 taken from the DECaLS survey. The size of the stamps is 2 arcminutes. North is up and east is left.}
\end{figure}

The final summary of the 27 
probable satellites of the Spindle galaxy is presented in Table 3.  Its columns contain: (1) galaxy name (name in {\it italic} for the galaxies observed by us with HST/ACS); (2) equatorial 
coordinates in degrees; (3) morphological type in de Vaucouleurs scale; (4) distance to the galaxy; (5)  method by which 
the distance is estimated: TRGB --- via the luminosity of the Tip of the Red Giant Branch, SBF ---
via fluctuation of surface brightness, TF --- from Tully-Fisher relation, NAM --- by the radial velocity of galaxy in 
the model of Numerical Action Method (Shaya et al.  2017), mem --- by implied membership in the group; (6) radial velocity 
of the galaxy relative to the Local Group centroid taken from NASA Extragalactic Database (=NED, http://ned.ipac.caltech.edu/); 
(7) integral luminosity of the galaxy in K-band from Updated Nearby Galaxy Catalog (= UNGC, http://www.sao.ru/lv/lvgdb) 
or from Carlsten et al. (2021a); (8) angular separation of the satellite from NGC~3115; (9) linear projected separation 
of the satellite under the assumption that it is at the same distance as the host galaxy; (10) the radial velocity 
difference between the satellite and the central galaxy; (11) orbital mass estimate of the total mass of NGC~3115 defined
 below. At the end of the Table, under the line, there are 4 galaxies with radial velocities close to that of NGC~3115, but
which are outside the "zero-velocity radius"  $R_0\simeq1.3$ Mpc for the NGC~3115 group. The galaxies are ranked by angular 
distance from NGC~3115.
  
  \begin{table}
 \caption{New probable satellites of NGC 3115.}
\begin{tabular}{cccclc} \hline

  Name      &   RA (J2000.0) DEC    &   $a\arcmin$    & b/a &     T  &      B  \\
\hline  
(1)&(2)&(3)&(4)&(5)&(6)\\
\hline
 Dw0956-0523&  09 56 55.6 -05 23 32 &   0.38  & 0.95&   dSph &  19.3 \\ 
 Dw0958-0834&  09 58 05.5 -08 34 36 &   0.37  & 0.73&   dSph &  19.6 \\
 Dw0958-0610&  09 58 35.3 -06 09 58 &   0.38  & 0.81&   dSph &  19.5 \\
 Dw1006-0828&  10 06 35.8 -08 27 56 &  0.16   &0.80 &  dIrr  & 21.6 \\
 Dw1007-0835&  10 07 15.4 -08 35 11 &  0.30   &0.33 &  dTr   & 20.3 \\
\hline
\end{tabular}
\end{table}  

\begin{figure}
    \centering
    \includegraphics[width=\textwidth]{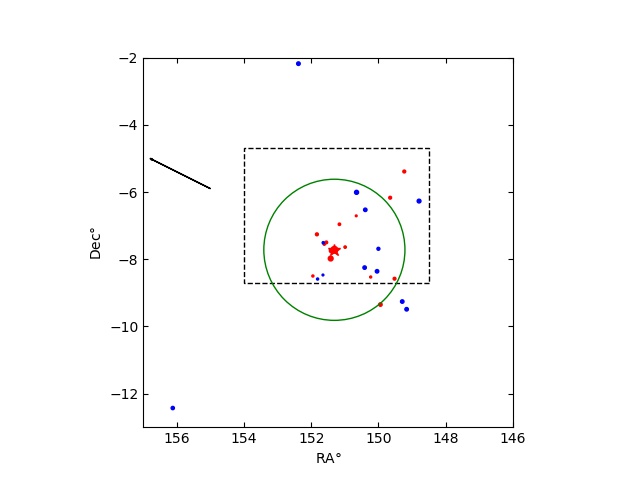}
    \caption{The distribution of companions around the Spindle galaxy (NGC~3115) in equatorial coordinates. 
The early-type galaxies and late-type galaxies are indicated by red and blue symbols, respectively. The green 
ring corresponds to the virial radius of 365 kpc around the NGC~3115. The rectangular perimeter delineates 
our viewing area on the DESI Lagacy Imaging Surveys. The arrow marks the direction towards the neighboring group around NGC~3521.}
\end{figure}
  
  Distribution of 27 supposed satellites of the Spindle galaxy is presented in Fig.4 in equatorial coordinates.
The galaxies of early types ( T $<$ 0) and late types ( T $>$ 0) are indicated by red and blue symbols, respectively.
The size of the circles is proportional to the logarithm of the galaxy's K-luminosity. The virial region with a radius 
$R_{vir}$ = 365 kpc = 2.05 deg, corresponding to the virial mass of $ 4.9 10^{12} M_{\odot} $ (see below), is outlined by a green ring.  The dotted rectangle shows the area of our view. Its 
southern perimeter is conditioned by the existing border of Legacy survey. The arrow points towards nearby bright 
galaxy NGC~3521. The panorama of dwarf retinue around NGC~3115 reveals several features:
 
 \begin{itemize}
     \item Almost half of the virial zone to the north-east of NGC~3115 is free of dwarf satellites down to the absolute magnitude $M_B$ = --10 mag.  
The reason for this asymmetry can be caused by the presence in this group of two flat (or linear) structures, to which the dwarf members are 
concentrated. This effect takes place in the Local Group and in other neighboring groups (Pawlowski et al. 2013; Ibata et al. 2014; Libeskind et al. 2015; Mueller et al. 2018).
     
     \item Distribution of the assumed satellites shows a tendency to group them into multiple subsystems at a scale of about 50 kpc. An example 
of such crowding in the Local Group is the pair of dwarfs NGC~147 and NGC~185. Radial velocity measurements of members of these substructures 
could clarify whether they are physically binded systems or due to a random projection effect. 
     
     \item The distribution of dwarf galaxies along the radius of the group shows signs of morphological segregation: dSphs are more common 
near the host galaxy than dIrrs. This is the well-known effect seen in the Local Group and other nearby groups (Mateo, 1998 and references therein).
However, there are several dSphs at a projected separation of 300--500 kpc, which indirectly indicates a 
significant extent of dark halo of Spindle galaxy.
     
 \end{itemize}
 
 The average projected separation of the 27 supposed satellites is $\langle R_p\rangle = 302\pm53$ kpc, and the 
rms difference in radial velocities of 10 satellites relative to NGC~3115 is 
$\langle \Delta V^2\rangle^{1/2} = 105$ km s$^{-1}$.  Following 
Karachentsev \& Kudrya (2014), we determined the orbital mass of the host Spindle galaxy as

   $$ M_T = (16/\pi) G^{-1} \langle \Delta V^2 R_p\rangle ,$$

where $G$ is the gravitation constant.  It is assumed here that the faint satellites of the central galaxy move 
along arbitrarily oriented Keplerian orbits with an average orbital eccentricity $e\simeq0.7$. Individual estimates 
of $M_T$ for
each satellite with measured radial velocity are indicated in the last column of Table 3.  The average orbital 
estimate of the total mass of NGC~3115 is $(4.89\pm1.48) 10^{12} M_{\odot}$, and the average ratio of the total 
mass
to the luminosity is $M_T/L_K = (50.0\pm15.1) M_{\odot}/L_{\odot}$, which is almost two times higher than the 
analogous ratio for the Milky Way, $(23\pm8) M_{\odot}/L_{\odot}$, and M~31, $(31\pm6) M_{\odot}/L_{\odot}$ 
(Karachentsev \& Kashibadze 2021).

  \section{A neighboring group of galaxies around NGC~3521.}
  As can be seen from the last row of Table 3, a bright ( B = 9.8 mag) neighboring Sbc galaxy NGC~3521 with a 
radial velocity  $V_h = 737$ km s$^{-1}$ is located at a projected separation of 17 deg or 3.0 Mpc from NGC~3115.
The radial distances of NGC~3521 (10.7 Mpc) and NGC~3115 (10.2 Mpc) practically coincide with each other within 
their measurement errors. According to Karachentsev \& Kashibadze (2021), NGC~3521 has 6 assumed satellites, of 
which radial velocities were measured for three.  We searched for new dwarf satellites using DECaLS data (Dey et 
al. 2019). In an area of 4 x 4 degrees (i.e. 750 x 750 kpc), we found four new likely satellites of low surface 
brightness.  Reproductions of their images from DECaLS survey 2' x 2' in size are shown in Fig.5.

\begin{figure}
    \centering
    \includegraphics[width=0.4\textwidth]{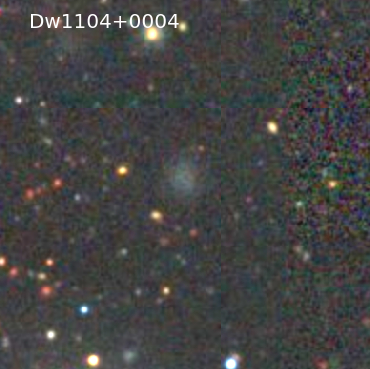}
    \includegraphics[width=0.4\textwidth]{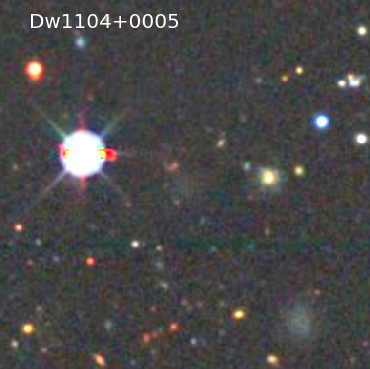}
    \includegraphics[width=0.4\textwidth]{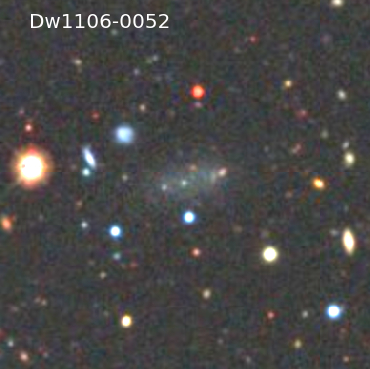}
    \includegraphics[width=0.4\textwidth]{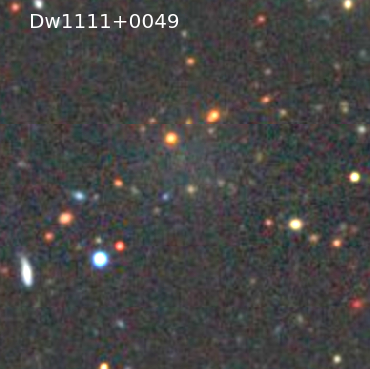}
    \caption{Color-composed images of four new probable companions of NGC~3521 taken from DECaLS survey. The size of the images is 2 arcminutes. North is up and east is left.}
\end{figure}

A general summary of data on 11 members of the NGC~3521 group is presented in Table 4, the structure of which is 
similar to that of Table 3.
  To date, none of the proposed satellites of NGC~3521 have had distances estimated. 
 The average projected separation of the satellites, $\langle R_p\rangle = 160\pm38$ kpc, and their rms 
velocity relative to NGC~3521,  $\langle\Delta V^2\rangle^{1/2} = 55$ km s$^{-1}$, turn out to be significantly 
less than that of the NGC~3115 satellites. The average estimate of the total mass 
$M_T = (0.90\pm0.42) 10^{12} M_{\odot}$ and the ratio $M_T/L_K = (7\pm3) M_{\odot}/L_{\odot}$ indicate a shallow 
potential well of this galaxy. The rotation curve of NGC~3521, which decreases with distance at periphery 
(Casertano \& van Gorkom 1991), also indicates a low mass of the dark halo of this spiral galaxy. 

  According to Tully (2015), the virial mass $M_T$ and virial radius of a group of galaxies $R_{vir}$ are connected
 by the empirical relation

         $$(R_{vir}/215\,\, {\rm kpc}) = (M_T/10^{12} M_{\odot})^{1/3}.$$

Applying it to the orbital mass estimates of NGC~3115 and NGC~3521, we obtain for them the expected virial radii 
of 365 kpc with a 1-$\sigma$ confidence interval of $[324-398]$ kpc and 208 kpc with a 1-$\sigma$ interval of 
$[168-232]$ kpc, respectively. These values are in good agreement with the average spatial separation of their 
satellites, $\langle R\rangle = (4/\pi) \langle R_p\rangle$, which is $(385\pm68$) kpc for NGC~3115 and ($204\pm48$) 
kpc for NGC~3521. Therefore, the average size of the suite of satellites around their central galaxy is a good 
indicator of the virial radius and the dark halo mass of the group. Following the relation by Tully (2015), a 
significant difference in the observed sizes of suites of these galaxies, 
$\langle R_p\rangle_{\rm N3115}/\langle R_p\rangle_{\rm N3521} \sim 1.9$, corresponds to the 
ratio of their virial mass  $\sim6$, which is close to the obtained ratio of the virial masses of these groups, $\sim 5.4$.
Note that the luminosities of the two parent galaxies are nearly the same: 10.99 dex vs. 11.09 dex.

   Both these groups, together with the group around NGC~2784 (D= 9.82 Mpc), are likely part of a diffuse elongated structure
"Leo Spur" and "Antlia cloud" (Tully, 1988), that is located near the rich Leo-I galaxy group having D= 11.3 Mpc.
\begin{table*}
\singlespace
\centering
\caption{Neighborhood of the galaxy NGC 3115.}
\begin{tabular}{lcrrrlcrrrr}\hline
Name     &   RA(2000)DEC &  T &    $D$   &  meth & $V_{LG}$ & $\log L_K$ & $r_p$  &  $R_p$  &  $\Delta V$  & $M_{orb}$  \\
  &    $^{\circ}$ $^{\circ}$   &    &   Mpc  &       & km\,s$^{-1}$ & $L_{\odot}$  & $^{\circ}$ &  kpc  & km\,s$^{-1}$ & $10^{12}$ \\
\hline
(1)&(2)&(3)&(4)&(5)&(6)&(7)&(8)&(9)&(10)&(11)\\ \hline
NGC 3115      & 151.31$-$07.72  &$-$1 &  10.2  &  TRGB,a) &  439 & 10.99  & 0.00 &    0  &   0  &  0.00 \\
KDG 65        &  151.40$-$07.75 &$-$3 &  10.2  &  mem  &  479 &  8.43  & 0.09 &   16  &  40  &  0.03 \\
KKSG 18       &  151.42$-$07.98 &$-$1 &  10.2  &  mem  &  456 &  9.31  & 0.28 &   50  &  17  &  0.02\\
dw1006-0730   &  151.55$-$07.50 &$-$2 &  10.2  &  mem  &  $-$  &  6.96  & 0.33 &   59  &  $-$  &  $-$  \\
dw1004-0737   &  150.99$-$07.64 &$-$2 &   7.8  &  SBF,b)  &  $-$  &  6.57  & 0.33 &   59  &  $-$  &  $-$  \\
dw1006-0732   &  151.61$-$07.55 &$-$2 &  10.2  &  mem  &  $-$  &  6.53  & 0.34 &   61  &  $-$  &  $-$  \\
dw1006-0730-n2&  151.64$-$07.51 & 10 &  10.2  &  mem  &  $-$  &  6.41  & 0.39 &   69  &  $-$  &  $-$  \\
dw1007-0715   &  151.83$-$07.26 &$-$2 &  10.2  &  mem  &  $-$  &  7.25  & 0.69 &  123  &  $-$  &  $-$  \\
dw1004-0657   &  151.16$-$06.96 &$-$2 &   8.8  &  SBF,b)  &  $-$  &  6.63  & 0.77 &  137  &  $-$  &  $-$  \\
Dw1006-0828   &  151.65$-$08.47 & 10 &  10.2  &  mem  &  $-$  &  5.74  & 0.82 &  147  &  $-$  &  $-$     \\
Dw1007-0835   &  151.81$-$08.59 & 10 &  10.2  &  mem  &  $-$  &  6.17  & 1.00 &  178  &  $-$  &  $-$     \\
dw1007-0830   &  151.95$-$08.50 &$-$2 &  10.3  &  SBF,b)  &  $-$  &  6.12  & 1.01 &  180  &  $-$  &  $-$    \\
{\it KKSG 17}       &  150.41$-$08.25 & 10 &   9.95 &  TRGB,c) &  203 &  7.87  & 1.04 &  185  &-236  & 12.16  \\
dw1002-0642   &  150.66$-$06.71 &$-$2 &  10.2  &  mem  &  $-$  &  5.66  & 1.20 &  214  &  $-$  &  $-$    \\
dw1000-0741   &  150.00$-$07.69 & 10 &  10.2  &  mem  &  $-$  &  7.30  & 1.31 &  233  &  $-$  &  $-$    \\
dw1000-0831   &  150.23$-$08.53 &$-$2 &  10.2  &  mem  &  $-$  &  6.06  & 1.35 &  240  &  $-$  &  $-$    \\
dw1000-0821   &  150.04$-$08.36 &  9 &  10.2  &  mem  &  $-$  &  7.91  & 1.42 &  253  &  $-$  &  $-$    \\
MCG-01-26-009 &  150.39$-$06.53 & 10 &  10.2  &  mem  &  510 &  7.87  & 1.50 &  267  &  71  &  1.59  \\
{\it UGCA 193}      &  150.65$-$06.01 &  7 &  10.00 &  TRGB,c) &  427 &  8.51  & 1.83 &  326  &$-$12  &  0.05  \\
Dw0958-0834   &  149.52$-$08.58 &$-$2 &  10.2  &  mem  &  $-$  &  7.09  & 1.99 &  354  &  $-$  &  $-$    \\
KKSG 16       &  149.94$-$09.35 &$-$3 &  10.2  &  mem  &  $-$  &  7.89  & 2.13 &  379  &  $-$  &  $-$   \\
Dw0958-0610   &  149.65$-$06.17 &$-$2 &  10.2  &  mem  &  $-$  &  7.13  & 2.27 &  404  &  $-$  &  $-$   \\
{\it PGC 154449}    &  149.29$-$09.26 &  9 &  10.13 &  TRGB,c) &  295 &  7.92  & 2.54 &  452  &-144  & 11.06 \\
LV J0956-0929 &  149.16$-$09.49 &  9 &   9.38 &  TRGB,d) &  378 &  7.95  & 2.78 &  495  &$-$61  &  2.17 \\
KKSG 15       &  148.79$-$06.27 & 10 &  10.2  &  mem  &  554 &  8.28  & 2.91 &  518  & 115  &  8.08 \\
Dw0956-0523   &  149.23$-$05.39 &$-$2 &  10.2  &  mem  &  $-$  &  7.21  & 3.12 &  556  &  $-$  &  $-$   \\
{\it PGC 1099440}   &  152.38$-$02.18 &  9 &  10.39 &  TRGB,c) &  519 &  8.03  & 5.64 & 1004  &  80  &  7.58 \\
{\it KKSG 19}       &  156.12$-$12.43 & 10 &  10.84 &  TRGB,c) &  373 &  7.70  & 6.73 & 1198  &$-$66  &  6.16 \\
\hline
AGC 202137    &  155.41$+$00.90 & 10 &   8.59 &  NAM  &  495 &  7.21  & 9.55 & 1700  &  56  &  6.29 \\
PGC 4078844   &  143.84$-$13.81 &  9 &   9.06 &  TF,e)   &  533 &  7.60  & 9.64 & 1716  &  94  & 17.89 \\
UGC 5797      &  159.85$+$01.72 &  9 &  10.42 &  TRGB,d) &  511 &  8.63  &12.73 & 2266  &  72  & 13.86\\
NGC 3521      &  166.45$-$00.04 &  4 &  10.70 &  TF,e)   &  598 & 11.09  &16.98 & 3022  & 159  &(90.15)\\
\hline
\multicolumn{11}{l}{
Notes: a) Peacock et al.2015; b) Carlsten, personal communication; c) present paper;} \\
\multicolumn{11}{l}{d) Anand et al. 2021; e) Karachentsev et al. 2013.} \\  
\multicolumn{11}{l} {\footnotesize{Alternative names: KDG 65 = UGCA 200 = PGC 29299, KKSG 18 = MCG-01-26-021 = PGC 29300, KKSG 17 =}} \\
\multicolumn{11}{l} {\footnotesize{= MCG-01-26-011 = PGC 29033, MCG-01-26-009 = PGC 29038, UGCA 193 = PGC 29086, KKSG 16 = PGC 3097699,}} \\
\multicolumn{11}{l} {\footnotesize{PGC 154449 = 2MASX J09570887-0915487, LV J0956-0929 = PGC 4078671, KKSG 15 = PGC 1034827, PGC 1099440 =}} \\
\multicolumn{11}{l} {\footnotesize{= 2dFGRS-TGN218Z179, KKSG 19 = HIPASS J1024-12 = PGC 3097700.}} \\ 
\hline
\end{tabular}
\end{table*}

\begin{table}
\centering
   \caption{Probable satellites of the galaxy NGC 3521.}
\begin{tabular}{lcrrlcrrrrr}\hline
     Name      &   RA(2000)DEC  &  T &    $D$   &  meth & $V_{LG}$ & $\log L_K$ & $r_p$  &  $R_p$  &  $\Delta V$  & $M_{orb}$  \\
             &    $^{\circ}$   &    &   Mpc  &       & km\,s$^{-1}$ & $L_{\odot}$  & $^{\circ}$ &  kpc  & km\,s$^{-1}$ & $10^{12}$ \\ 
 \hline 
 (1)&(2)&(3)&(4)&(5)&(6)&(7)&(8)&(9)&(10)&(11)\\
 \hline

 NGC 3521     &    166.45$-$00.04 &        4   &    10.7   &    TF  &    598    &      11.09   &       0  &     0    &    0    &      0    \\
 N3521sat     &    166.42 +00.12 &     $-$1   &    10.7   &   mem  &    $-$    &       8.62   &    0.16  &     30   &   $-$   &     $-$  \\
 KKSG 20      &    166.17 +00.06 &       10   &    10.7   &   mem  &    636    &       7.63   &    0.30  &     56   &   38    &    0.10 \\
 Dw1104+0005  &    166.17 +00.09 &     $-$2   &    10.7   &   mem  &    $-$    &       6.58   &    0.31  &     58   &   $-$   &     $-$ \\
 Dw1104+0004  &    166.16 +00.08 &     $-$2   &    10.7   &   mem  &    $-$    &       7.02   &    0.32  &     59   &   $-$   &     $-$ \\
 N3521dwTBG   &    166.80$-$00.19 &     $-$2   &    10.7   &   mem  &    $-$    &       7.28   &    0.38  &      72  &    $-$  &      $-$\\
 Dw1106-0052  &    166.61$-$00.88 &       10   &    10.7   &   mem  &    $-$    &       6.67   &    0.86  &     161  &   $-$   &     $-$ \\
 dw1110+0037  &    167.62 +00.62 &       10   &    10.7   &   mem  &     669   &       7.62   &    1.34  &     252  &    71   &    1.50 \\  
 KKSG 22      &    166.53$-$01.45 &       10   &    10.7   &   mem  &    $-$    &       7.11   &    1.41  &     265  &    $-$  &      $-$\\
 Dw1111+0049  &    167.85 +00.83 &       10   &    10.7   &   mem  &    $-$    &       6.67   &    1.64  &     306  &    $-$  &      $-$\\
 UGC 6145     &    166.39$-$01.86 &       10   &    10.7   &   mem  &     546   &       7.83   &    1.82  &     341  & $-$52   &    1.09 \\
\hline
\multicolumn{11}{l} {Alternative names: KKSG 20 = PGC 135770, dw1110+0037 = SDSS J111029.56+003700.7.} \\
\hline
\end{tabular}
\end{table}
   
\section{Isolated luminous early-type galaxies in the Local Volume.}
  As noted above, in the Local Volume, limited by the distance $D < 11$ Mpc, there are 5 relatively isolated ETG 
galaxies with $L_K/L_{\odot}$ luminosities brighter than 10.5 dex.  Here we do not take into account several bright 
E and S0 galaxies in the central part of the Leo-I group (NGC~3379, NGC~3384, NGC~3489), which looks like a
small cluster with a united dark halo.  Data on the 5 detached ETG galaxies are presented in Table 5. The 
table columns contain: (1) galaxy name; (2) morphological type; (3,4) distance to the galaxy ( Mpc) and the 
method used to determine the distance; (5) number of known supposed satellites; (6) number of satellites with measured 
radial velocities; (7) number of satellites with accurate distances measured via TRGB method; (8) $L_K$
luminosity of ETG galaxy in the $L_{\odot}$ units; (9) the ratio of the total (orbital) mass-to-$L_K$-luminosity in 
units of $M_{\odot}/L_{\odot}$ according to Karachentsev \& Kashibadze (2021).
 
 As one can see, the nearest group around NGC~5128 (Cen~A) has been studied in much more detail than the other groups 
on the outskirts of the Local Volume. The number of  assumed satellites around NGC~4594 (Sombrero), NGC~3115, NGC~2784 
and NGC~1291 is quite representative ( n = 17-- 28), but only a small part of them have individual distance estimates 
and measured radial velocities.  The least reliable data are for galaxies in the vicinity of NGC~2784.  Five 
putative satellites of this galaxy: DDO~56, HIPASSJ0916-23b, ESO~497-004, NGC~2835 and DDO~62 are located at projected 
separations of $R_p = (220 - 460)$ kpc and they all have a positive difference in radial velocities with respect to 
NGC~2784. Therefore, the orbital estimate of the mass of this group is highly doubtful.  For the other four ETG 
galaxies, the average ratio of total mass-to-luminosity is $\langle M_T/L_K\rangle = 58\pm6 (M_{\odot}/L_{\odot}$).  
Taking, according to Lelli et al. (2016) the value $M_*/L_K = 0.7 M_{\odot}/L_{\odot}$ for the bulge population, we obtain for luminous ETG 
galaxies the ratio $\langle M_T/M_*\rangle = 83\pm9$.
 \begin{table}
   \caption{Isolated early-type galaxies in the Local Volume with $L_K/L_{\odot} > 10.5$ dex.}
\begin{tabular}{clrccrrcr}\hline
(1)&(2)&(3)&(4)&(5)&(6)&(7)&(8)&(9)\\ \hline

 Name        &      Type   &     $D$  &       meth &     $n$ &      $n_v$ &    $n_{\rm TRGB}$  &    $\log(L_K)$  &   $M_T/L_K$      \\
\hline
 NGC 5128    &    S0p      &   3.68  &    TRGB    &  62   &   34     &  43        &     10.89    &      60$\pm$18   \\
 NGC 4594    &    S0a      &   9.55  &    TRGB    &  27   &   15     &    3       &      11.32   &       74$\pm$24  \\
 NGC 3115    &    S0       &   10.2  &   TRGB     &  27   &   10     &    6       &      10.99   &       50$\pm$15  \\
 NGC 2784    &    S0       &   9.82  &    SBF     & 28    &    5:    &    0       &      10.80   &      145$\pm$50: \\   
 NGC 1291    &    SB0a     &  10.97  &   TRGB     & 17    &    2     &    1       &      10.97   &        47$\pm$4 \\
\hline 
\end{tabular}
\end{table}
 
  \section{Concluding remarks.}
  Relatively isolated E,S0 galaxies in the Local Volume with luminosity of $\log(L_K/L_{\odot}) > 10.5$ have the average luminosity 
of $\langle\log(L_K/L_{\odot})\rangle =10.99\pm0.09$ and the average total mass-to-luminosity ratio $\langle M_T/L_K\rangle =
58\pm6 (M_{\odot}/L_{\odot})$.  There are only 5 such galaxies out of 1200 galaxies in the Local Volume.  Despite the 
poor statistics, these parameters turn out to be quite representative for the population of massive ETG galaxies.
Thus, a sample of 26 isolated E,S0 galaxies of the northern sky from the KIG-catalog, which have faint companions with 
measured radial velocities, is characterized by the  values: $\langle\log(L_K/L_{\odot})\rangle = 11.01\pm0.06$ and 
$\langle M_T/L_K\rangle = 74\pm26 (M_{\odot}/L_{\odot})$ (Karachentseva et al. 2021a). For 60 isolated ETG galaxies from 
the 2MIG catalog, in which the radial velocities of satellites were measured, the median parameters are 
$\log(L_K/L_{\odot}) = 11.13$ and $M_T/L_K = 63 (M_{\odot}/L_{\odot})$ (Karachentseva et al. 2011).  Note that both 
catalogs of isolated galaxies, KIG and 2MIG, selected from optical and infrared (2MASS) sky surveys, cover significant 
volume up to redshift $z\sim0.02$.
  
  A similar analysis of data on 141 dwarfs with known radial velocities around isolated spiral KIG- galaxies gives an 
average ratio of $\langle M_T/L_K\rangle = 20.9\pm3.1 (M_{\odot}/L_{\odot})$ (Karachentseva et al. 2021b).  For isolated 
spiral galaxies of the 2MIG catalog, the median of this ratio via 154 satellites is 17 $M_{\odot}/L_{\odot}$ 
(Karachentseva et al. 2011). Part of this difference is obviously due to different $M_*/L_K$ ratios for the stellar 
population of the disk (0.5 $M_{\odot}/L_{\odot}$) and bulge (0.7 $M_{\odot}/L_{\odot}$) (Lelli et al. 2016). Generally, 
the available data on the motions of faint satellites around isolated high-luminosity 
galaxies definitely indicate a (2--3)-fold excess of dark matter around bulge-dominated galaxies compared to disk-dominated 
ones of the same stellar mass. This conclusion is consistent with the assertions: "we find that passive central galaxies 
have halos that are at least twice massive as those of star-forming objects of the same stellar mass" (Mandelbaum et al. 2016) 
and "the red galaxies occupy dark matter halos that are much more massive that those occupied by blue galaxies with the 
same stellar mass" (Bilicki et al. 2021), both obtained from the weak gravitational lensing data.

 Despite the limited statistics, there is also a noticeable tendency according to which the 
variation in the ratio of dark-to-light matter $(M_T/M_*)$ in late-type galaxies is much wider than in early-type galaxies. 
Moreover, some spiral galaxies with a decreasing rotation curve at the periphery (NGC~253, NGC~2683, NGC~2903, NGC~3521 
and NGC~5055) have an anomalously low ratio $M_T/L_K \sim 5 M_{\odot}/L_{\odot}$ (Karachentsev et al. 2021). The observed 
diversity of luminous galaxies of different morphological types in terms of the $M_T/L_K$ parameter has not yet found a 
clear theoretical interpretation.

\begin{acknowledgments}
  We are grateful to the anonymous referee for a prompt report that helped us improve the manuscript.
  This work is based on observations made with the NASA/ESA Hubble Space Telescope. Support for program SNAP--15922 
(PI R.B.Tully) was provided by NASA through a grant from the Space Telescope Science Institute, which operated by the 
Associations of Universities for Research in Astronomy, Incorporated, under NASA contract NASb5-26555.  
Part of the work was made using the facilities of the Big Telescope Alt-azimutal SAO RAS and supported under  the   Ministry of Science and Higher Education of the Russian Federation grant 075-15-2022-233 (13.MNPMU.21.0003).

We used also data 
obtained with the Dark Energy Camera Legacy Survey (DECaLS) founded by the U.S. Department of Energy, the  U.S. National 
Science Foundation.
\end{acknowledgments}

  \bigskip
 {\bf References}
 
Anand G.S., Rizzi L., Tully R.B., et al., 2021, AJ, 162, 80

Bilek M., Samurovic S., Renauld F., 2019, A \& A, 629L, 5

Bilicki M., Dvornik A., Hoekstra H., et al. 2021, A \& A, 653A, 82

Capaccioli M., Cappellaro E., Held E.V., Vierti M., 1993, A \& A, 274, 69

Carlsten S.G., Greene J.E., Greco J.P., et al. 2021a, ApJ, 922, 267

Carlsten S.G., Greene J.E., Beaton R.L., Greco J.P., 2021b, arXiv:2105.03440

Casertano S. \& van Gorkom J.H., 1991, AJ, 101, 1231

Colles M., Peterson B.A., Jackson C.A. et al. 2003, arXiv:astro-ph/0306581 (2dF)

Dey A., Schlegel D.J., Lang D., et al. 2019, AJ, 157, 168

Dolphin, A.E. 2000, PASP, 112, 1383

Dolphin, A.E. 2016, DOLPHOT: Stellar photometry, ascl:1608.013

Doyle M.T., Drinkwater M.J., Rohde D.J. et al. 2005, MNRAS, 361, 34

Guerou A., Emsellem E., Krajnovich D., et al. 2016, A \& A, 591A, 143

Huchtmeier W.K., Karachentsev I.D., Karachentseva V.E., 2001, A \& A, 377, 801

Ibata R.A., Ibata N.G., Lewis G.F., et al, 2014, ApJ, 784L, 6

Jones D.H., Read M.A., Saunders W., et al. 2009, MNRAS, 399, 683 (6dF)

Karachentsev I.D., Kashibadze O.G., 2021, AN, 342, 999

Karachentsev I.D., Tully R.B., Anand G.S., et al. 2021, AJ, 161, 205.

Karachentsev I.D., Makarova L.N., Tully R.B., et al. 2020, A \& A, 643A, 124

Karachentsev I.D., Kudrya Y.N., 2014, AJ, 148, 50

Karachentsev I.D., Makarov D.I., Kaisina E.I., 2013, AJ, 145, 101

Karachentsev I.D., Karachentseva V.E., Suchkov A.A., Grebel E.K., 2000, A \& AS, 145, 41

Karachentseva V.E., Karachentsev I.D., Melnyk O.V., 2011, Astrophys. Bull., 66, 389

Karachentseva V.E., Karachentsev I.D., Melnyk O.V., 2021a, Astrophys. Bull., 76, 124

Karachentseva V.E., Karachentsev I.D., Melnyk O.V., 2021b, Astrophys. Bull., 76, 341

Kourkchi E., Tully R.B., 2017, ApJ, 843, 16

Lapi A., Salucci P., Danese L., 2018, ApJ, 859, 2

Lelli F., McGaugh S.S., Schombert J.M., 2016, AJ, 152, 157

Libeskind N.I., Hoffman Y., Tully R.B., et al. 2015, MNRAS, 452, 1052

Makarov, D.I., Prugniel P., Terekhova N., et al, 2014, A\&A, 570A, 13

Makarov, D.I., Makarova, L.N., Rizzi, L. et al., 2006, AJ, 132, 2729

Mandelbaum R., Wang W., Zu Y., et al, 2016, MNRAS, 457, 3200

Mateo M., 1998, ARA \& A, 36, 435

Moster B.P., Naab T., White S.D.M., 2013, MNRAS, 428, 3121

Muller O., Pawlowski M.S., Jerjen H., Lelli F., 2018, Sci, 359, 534

Pawlowski M.S., Kroupa P., Jerjen H., 2013, MNRAS, 435, 2116

Peacock M.B., Strader J., Romanowsky A.J., Brodie J.P., 2015, ApJ, 800, 13

Peng E.W., Ford H.C., Freeman K.C., 2004, ApJ, 602, 685

Posti L. \& Fall S.M., 2021, A \& A, 649, 119

Rizzi, L., Tully, R.B., Makarov, D.I. et al. 2007, ApJ, 661, 815

Sales L.V., Wang W., White S.D.M., Navarro J.F., 2013, MNRAS, 428, 573

Santos-Santos I.M.E., Sales L.V., Fattahi A., Navarro J.F., 2021, arXiv:2111.01158

Schlafly, E.F., Finkbeiner, D.P. 2011, ApJ, 737, 103

Shaya E.J., Tully R.B., Hoffman Y., et al, 2017, ApJ, 850, 207

Tonry J.L., Dressler A., Blakeslee J.P., et al. 2001, ApJ, 546, 681

Tully R.B., 2015, AJ, 149, 54

Tully R.B., 1988, Nearby Galaxies Catalog, Cambridge University Press, Cambridge

Vorontsov-Veliaminov B.A., Arkhipova V.P., 1963, Morphological Catalog of Galaxies, part III, Moscow State University, Moscow

Wechsler R.H. \& Tinker L.J., 2018, ARA\&A, 56, 435

Wegner G., Bernardi M., Willmer C.N.A., et al. 2003, AJ, 126, 2268

Wu, P.-F., Tully, R.B., Rizzi, L. et al. 2014, AJ, 148, 7

\end{document}